\documentclass[12pt]{article}
\usepackage[dvips]{graphicx}
\usepackage{amsmath,amssymb,amsthm,fullpage}
\usepackage{mathtext} % если нужны русские буквы в формулах (не обязате
\usepackage{inputenc} % кодировка - можно использовать [koi8-r][cp866][cp1251]
\usepackage[T2A]{fontenc}  % внутренняя T2A кодировка TeX
\usepackage[english]{babel}   % включение переносов
\usepackage{cite}
\renewcommand\le{\leqslant}
\renewcommand\ge{\geqslant}

\newcommand{\Nset}{\mathbb N}

\newcommand{\Zset}{\mathbb Z}
\newcommand{\Rset}{\mathbb R}
\newcommand{\Cset}{\mathbb C}

\newcommand{\Mgoth}{{\mathfrak M}}

\def\Ai{{\rm Ai}}
\def\Bi{{\rm Bi}}

\newcommand{\rme}{\mathrm{e}}
\newcommand{\rmi}{\mathrm{i}}
\newcommand{\rmd}{\mathrm{d}}

\renewcommand{\Re}{\mathop{\rm Re}\nolimits}
\renewcommand{\Im}{\mathop{\rm Im}\nolimits}
\newcommand{\sgn}{\mathop{\rm sgn}\nolimits}

\allowdisplaybreaks

\title{Mellin transform of quartic products\\ of shifted Airy functions}

\author{E.G. Abramochkin and E.V. Razueva\\ 
Coherent Optics Lab, Lebedev Physical Institute,\\ 
Samara, 443011, Russia}
\date{\today}

\begin{document}

\maketitle
\begin{abstract}
The Mellin transform of quartic products of shifted Airy functions is 
evaluated in a closed form. Some particular cases expressed in terms of 
the logarithm function and complete elliptic integrals special values are 
presented.
\end{abstract}

%\ams{33C10, 33C05, 44A20}

%\noindent{\it Keywords\/}: Mellin transform, Airy functions, hypergeometric function

%\submitto{\jpa}

\section{Introduction}\label{sec1}
%%%%%%%%%%%%%%%%%%%%%%%%%%%%%%%%%%%%%%%%%%%%%%%%%%%%%%%%%%%%%%%%%%%%%%%%%%%%

This article is inspired by the Laurenzi integral \cite{Laurenzi} (see 
also \cite[Eq.9.11.18]{DLMF})
\begin{equation}
\int_0^\infty \Ai^4(x)\,\rmd x=\frac{\ln 3}{24\pi^2}.
\label{eq:Laurenzi}
\end{equation}
As far as we know, it was the first case when the integral of Airy function 
power results in a logarithmic function value.

Products of two Airy functions and their integral transforms are investigated 
in many mathematical and physical papers (see, for example, \cite{Moyer, 
Berry1979, Berry1980, Schwinger, Varlamov2011, Berry2013})
while products of three or more Airy functions meet quite rarely.
Besides the Laurenzi paper [1], these products are discussed in \cite{Aspnes1, 
Reid2, Reid3, ValleeSoares1997, AiryBook, Masomi}.

Recently, a two-dimensional light field that is a product of three Airy 
functions with linear arguments has been proposed and invesigated in 
\cite{tAi}. It has been shown that the Fourier image of this {\it three-Airy 
beam} has a radially symmetric intensity with a super-Gaussian decrease. 
Three-Airy beams, Airy wavelets in two dimensions and light fields based on 
cubic and quartic products of Airy functions have been investigated in 
\cite{Torre2014, Torre2015}.

Below we consider the quartic products only. The paper is 
organized as follows. In section~2 we evaluate the Mellin transform of the 
function $\Ai^4(c+x)$, where $c$ is an arbitrary complex parameter. In 
section~3 similar results for the functions $\Ai^3(c+x)\Bi(c+x)$ and 
$\Ai^2(c+x)\Bi^2(c+x)$ are obtained. Some examples including the case when 
the final expression may be written in terms of complete elliptic integrals 
are also given there. In the last section we shortly discuss the Mellin 
transform of other products of Airy functions and application of the above 
results to evaluation of the Gauss hypergeometric function special values.

\section{The Mellin transform of $\Ai^4(c+x)$}\label{sec2}
%%%%%%%%%%%%%%%%%%%%%%%%%%%%%%%%%%%%%%%%%%%%%%%%%%%%%%%%%%%%%%%%%%%%%%%%%%%%

Let us consider the problem of evaluation of the integral
\begin{equation}
\Mgoth(\alpha,c)=\Mgoth[\Ai^4(c+x)](\alpha)
=\int_0^\infty x^{\alpha-1}\Ai^4(c+x)\,\rmd x,
\label{eq:Moments_Ai4_c}
\end{equation}
where $\alpha>0$. For simplicity we assume that $c\in\Rset$ while, of course, 
due to the asymptotic behaviour of the Airy function the final result will be 
valid for any $c\in\Cset$.

All the following manipulations are quite straightforward and based on 
traditional methods used in calculus and complex analysis, such as series 
expansions, contour integration and analytic continuation.

We start with two Aspnes integrals~\cite{Aspnes1} which may be written in 
the united form:
\begin{equation}
\Ai^2(x)+\rmi\Ai(x)\Bi(x)=\frac{1}{2\pi^{3/2}}\!
\int_0^\infty \exp\Bigl(\rmi\Bigl[\frac{t^3}{12}+xt
+\frac{\pi}{4}\Bigr]\Bigr)\frac{\rmd t}{\sqrt t}\qquad (x\in\Rset).
\label{eq:Aspnes}
\end{equation}
Then we change $x\to c+x$ in (\ref{eq:Aspnes}), separate its real part:
\begin{equation}
\Ai^2(c+x)=\frac{1}{4\pi^{3/2}}\sum_{\epsilon=\pm 1}\int_0^\infty 
\exp\Bigl(\rmi\epsilon\Bigl[\frac{t^3}{12}+(c+x)t+\frac{\pi}{4}\Bigr]\Bigr)
\frac{\rmd t}{\sqrt t}\,,
\label{eq:Ai^2(c+x)_Int}
\end{equation}
and substitute two copies of (\ref{eq:Ai^2(c+x)_Int}) into 
(\ref{eq:Moments_Ai4_c}):
\begin{align}
\Mgoth(\alpha,c)&=\frac{1}{16\pi^3}\int_0^\infty \!x^{\alpha-1}\,\rmd x
\nonumber\\
&{}\times 
\sum_{\delta=\pm 1}\sum_{\epsilon=\pm 1}\int_0^\infty \!\!\!\int_0^\infty
\exp\Bigl(\rmi\delta\Bigl[\frac{s^3}{12}+(c+x)s+\frac{\pi}{4}\Bigr]
+\rmi\epsilon\Bigl[\frac{t^3}{12}+(c+x)t+\frac{\pi}{4}\Bigr]\Bigr)
\frac{\rmd s\,\rmd t}{\sqrt{st}}
\nonumber\\
&{}=\frac{1}{8\pi^3}\Re\int_0^\infty \!x^{\alpha-1}\,\rmd x\biggl\{
\int_0^\infty \!\!\!\int_0^\infty
\exp\Bigl(\rmi\Bigl[\frac{s^3+t^3}{12}+(c+x)(s+t)+\frac{\pi}{2}\Bigr]\Bigr)
\frac{\rmd s\,\rmd t}{\sqrt{st}}
\nonumber\\
&\hspace{55mm}{}+\int_0^\infty \!\!\!\int_0^\infty
\exp\Bigl(\rmi\Bigl[\frac{s^3-t^3}{12}+(c+x)(s-t)\Bigr]\Bigr)
\frac{\rmd s\,\rmd t}{\sqrt{st}}\biggr\}\,.
\nonumber
\end{align}
Here the first double integral corresponds to the terms with 
$(\delta,\epsilon)=(+1,+1)\cup(-1,-1)$, and the second one --- to the terms 
with $(\delta,\epsilon)=(+1,-1)\cup(-1,+1)$.

Using the known formula
$$
\int_0^\infty \exp(\rmi\lambda x)\,x^{\alpha-1}\,\rmd x
=\frac{\Gamma(\alpha)}{|\lambda|^\alpha}
\exp\Bigl(\frac{\pi\rmi\alpha}{2}\sgn\lambda\Bigr)
\qquad (\lambda\in\Rset,\; \lambda\ne 0),
$$
we integrate over $x$ and obtain
\begin{align}
\Mgoth(\alpha,c)&=\frac{\Gamma(\alpha)}{8\pi^3}
\Re\biggl\{\int_0^\infty \!\!\!\int_0^\infty
\exp\Bigl(\rmi\Bigl[\frac{s^3+t^3}{12}+c(s+t)
+\frac{\pi(\alpha+1)}{2}\Bigr]\Bigr)
\frac{\rmd s\,\rmd t}{\sqrt{st}\,(s+t)^\alpha}
\nonumber\\
&\hspace{20mm}{}+\int_0^\infty \!\!\!\int_0^\infty
\exp\Bigl(\rmi\Bigl[\frac{s^3-t^3}{12}+c(s-t)
+\frac{\pi\alpha}{2}\sgn(s-t)\Bigr]\Bigr)
\frac{\rmd s\,\rmd t}{\sqrt{st}\,|s-t|^\alpha}\biggr\}
\nonumber\\
&{}=\frac{\Gamma(\alpha)}{8\pi^3}
\Re\bigl\{\rme^{\pi\rmi(\alpha+1)/2}I_1+I_2\bigr\}.
\label{eq:Mgoth1}
\end{align}
To insure the convergence of the integrals we restrict our study to the 
case of small $\alpha$'s, namely, $\alpha\in(0,1)$. For other values of 
$\alpha$ the final result may be established by analytic continuation (see 
\cite{Watson_ENG, Nikiforov_ENG} for details).

We consider first $I_1$, then $I_2$.

To evaluate the integral $I_1$ we use the change of variables $u=s+t$, $v=st$. 
Then
\begin{gather}
s^3+t^3=(s+t)(s^2-st+t^2)=u(u^2-3v),
\nonumber\\
\rmd u\,\rmd v=\Bigl|\frac{\partial(u,v)}{\partial(s,t)}\Bigr|\,\rmd s\,\rmd t
=|s-t|\,\rmd s\,\rmd t=\sqrt{u^2-4v}\,\rmd s\,\rmd t,
\nonumber
\end{gather}
and the first quadrant of $(s,t)$-plane is mapped twice onto the region 
$\bigl\{ v\ge 0,\; u\ge 2\sqrt{v}\bigr\}$ of $(u,v)$-plane. It is easy 
to verify images of half-lines $\bigl\{s\ge 0,\; t\ge 0,\; 
{t=s\tan\theta}\bigr\}$ for each $\theta\in(0,\pi/2)$ under the mapping 
$(s,t)\to (u,v)$. Then
\begin{align}
I_1&=2\!\int_0^\infty \frac{\rmd u}{u^\alpha}\int_0^{u^2/4} 
\exp\Bigl(\frac{\rmi u(u^2-3v)}{12}+\rmi cu\Bigr)
\frac{\rmd v}{\sqrt{v(u^2-4v)}}
\nonumber\\
&{}=\int_0^\infty \frac{\rmd u}{u^\alpha}\int_0^1 
\exp\Bigl(\frac{\rmi u^3}{12}\Bigl[1-\frac{3\tau}{4}\Bigr]+\rmi cu\Bigr)\,
\frac{\rmd\tau}{\sqrt{\tau(1-\tau)}}\,,
\nonumber
\end{align}
where the last equality has been found by using the change of variable 
$v=(u^2/4)\tau$.

Applying contour integration, we transform the integral over $u$ into an 
absolutely convergent one,
$$
I_1=\exp\Bigl(\frac{\pi\rmi}{6}(1-\alpha)\Bigr)
\int_0^1 \!\frac{\rmd\tau}{\sqrt{\tau(1-\tau)}}\int_0^\infty 
\exp\Bigl(-\frac{u^3}{12}\Bigl[1-\frac{3\tau}{4}\Bigr]
+\rme^{2\pi\rmi/3}cu\Bigr)\frac{\rmd u}{u^\alpha}\,,
$$
and expand the factor $\exp(\rme^{2\pi\rmi/3}cu)$ in the Taylor series for 
the following term-by-term integration:
\begin{align}
I_1&{}=\exp\Bigl(\frac{\pi\rmi}{6}(1-\alpha)\Bigr)
\sum_{n=0}^\infty \frac{(\rme^{2\pi\rmi/3}c)^n}{n!}
\int_0^1 \!\frac{\rmd\tau}{\sqrt{\tau(1-\tau)}}
\int_0^\infty \exp\Bigl(-\frac{u^3}{12}\Bigl[1-\frac{3\tau}{4}\Bigr]\Bigr)
\,u^{n-\alpha}\,\rmd u
\nonumber\\
&{}=\frac13\exp\Bigl(\frac{\pi\rmi}{6}(1-\alpha)\Bigr)
\sum_{n=0}^\infty \,\Gamma\Bigl(\frac{n+1-\alpha}{3}\Bigr)
12^{(n+1-\alpha)/3}\frac{(\rme^{2\pi\rmi/3}c)^n}{n!}
\int_0^1 \!\frac{\tau^{-1/2}(1-\tau)^{-1/2}}
{\bigl(1-\tfrac34\tau\bigr)^{(n+1-\alpha)/3}}\,\rmd\tau
\nonumber\\
&{}=\frac{\pi}{3}\sum_{n=0}^\infty \,\Gamma\Bigl(\frac{n+1-\alpha}{3}\Bigr)
12^{(n+1-\alpha)/3}\exp\Bigl(\frac{\pi\rmi}{6}(4n+1-\alpha)\Bigr)
\frac{c^n}{n!}\cdot
{}_2F_1\Bigl(\frac{n+1-\alpha}{3},\frac12;\,1\,\Bigl|\,\frac34\Bigr).
\nonumber
\end{align}
Here we used the known formula
$$
\int_0^\infty \exp(-\lambda x^n)\,x^{\alpha-1}\,\rmd x
=\frac{1}{n\lambda^{\alpha/n}}\Gamma\Bigl(\frac{\alpha}{n}\Bigr)
\qquad (\Re\lambda>0,\; \Re(\lambda^{\alpha/n})>0)
$$
and the integral representation of the Gauss hypergeometric function 
\cite[Eq.(7.2.1.2)]{Prudnikov3_ENG}:
\begin{equation}
{}_2F_1(a,b;\,c\,|\,z)=\frac{\Gamma(c)}{\Gamma(b)\Gamma(c-b)}
\int_0^1 \frac{t^{b-1}(1-t)^{c-b-1}}{(1-zt)^a}\,\rmd t.
\label{eq:HyperG_2F1_Int}
\end{equation}

As result, the contribution of $I_1$ to $\Mgoth(\alpha,c)$ is
\begin{align}
\frac{\Gamma(\alpha)}{8\pi^3}
\Re\bigl\{\rme^{\pi\rmi(\alpha+1)/2}I_1\bigr\}
=\frac{\Gamma(\alpha)}{24\pi^2}
\sum_{n=0}^\infty \,\Gamma\Bigl(\frac{n+1-\alpha}{3}\Bigr)\,
12^{(n+1-\alpha)/3}\cos\Bigl(\frac{\pi}{3}(2n+2+\alpha)\Bigr)\frac{c^n}{n!}
\nonumber\\
{}\times{}_2F_1\Bigl(\frac{n+1-\alpha}{3},\frac12;\,1\,\Bigl|\,\frac34\Bigr).
\nonumber
\end{align}

To evaluate the integral $I_2$ we use the change of variables $u=s-t$, $v=st$. 
Then
$$
s^3-t^3=u(u^2+3v),\qquad 
\rmd u\,\rmd v=|s+t|\,\rmd s\,\rmd t=\sqrt{u^2+4v}\,\rmd s\,\rmd t,
$$
and the first quadrant of $(s,t)$-plane is mapped onto the upper half-plane 
$\{u\in\Rset,\,v\ge 0\}$ of $(u,v)$-plane because half-lines $\bigl\{s\ge 0,\; 
t\ge 0,\; {t=s\tan\theta}\bigr\}$ are mapped onto parabolas 
$\{u=s(1-\tan\theta),\, v=s^2\tan\theta\}$. Then
\begin{align}
I_2&=\int_\Rset \rmd u\int_0^\infty 
\exp\Bigl(\frac{\rmi u(u^2+3v)}{12}+\rmi cu+\frac{\pi\rmi\alpha}{2}\sgn u\Bigr)
\frac{\rmd v}{\sqrt{v(u^2+4v)}\,|u|^\alpha}={}
\nonumber\\
&{}=2\Re\int_0^\infty \frac{\rmd u}{u^\alpha}\int_0^\infty 
\exp\Bigl(\frac{\rmi u(u^2+3v)}{12}+\rmi cu+\frac{\pi\rmi\alpha}{2}\Bigr)
\frac{\rmd v}{\sqrt{v(u^2+4v)}}\,.
\nonumber
\end{align}
The following steps are similar to that for $I_1$:
\begin{align}
I_2&=\Re\int_0^\infty \frac{\rmd u}{u^\alpha}\int_0^\infty 
\exp\Bigl(\frac{\rmi u^3}{12}\Bigl[1+\frac{3\tau}{4}\Bigr]
+\rmi cu+\frac{\pi\rmi\alpha}{2}\Bigr)\frac{\rmd\tau}{\sqrt{\tau(1+\tau)}}
\nonumber\\
&{}=\Re\biggl\{\exp\Bigl(\frac{\pi\rmi}{6}(2\alpha+1)\Bigr)
\int_0^\infty \!\!\frac{\rmd\tau}{\sqrt{\tau(1+\tau)}}\,
\int_0^\infty \exp\Bigl(-\frac{u^3}{12}\Bigl[1+\frac{3\tau}{4}\Bigr]
+\rme^{2\pi\rmi/3}cu\Bigr)\frac{\rmd u}{u^\alpha}\biggr\}
\nonumber\\
&{}=\sum_{n=0}^\infty \cos\Bigl(\frac{\pi}{6}(4n+1+2\alpha)\Bigr)\frac{c^n}{n!}
\int_0^\infty \!\!\frac{\rmd\tau}{\sqrt{\tau(1+\tau)}}\,
\int_0^\infty \exp\Bigl(-\frac{u^3}{12}\Bigl[1+\frac{3\tau}{4}\Bigr]\Bigr)
u^{n-\alpha}\,\rmd u
\nonumber\\
&{}=\frac13\sum_{n=0}^\infty \,\Gamma\Bigl(\frac{n+1-\alpha}{3}\Bigr)
12^{(n+1-\alpha)/3}\cos\Bigl(\frac{\pi}{6}(4n+1+2\alpha)\Bigr)\frac{c^n}{n!}
\int_0^\infty \!\!\frac{\tau^{-1/2}(1+\tau)^{-1/2}}
{\bigl(1+\tfrac34\tau\bigr)^{(n+1-\alpha)/3}}\,\rmd\tau\,.
\nonumber
\end{align}
The last integral is reduced to the Gauss hypergeometric function by using 
the change of variable $\tau=t/(1-t)$ and formula 
(\ref{eq:HyperG_2F1_Int}):
\begin{align}
\int_0^\infty \!\!\frac{\tau^{-1/2}(1+\tau)^{-1/2}}
{\bigl(1+\tfrac34\tau\bigr)^{(n+1-\alpha)/3}}\,\rmd\tau
&=\int_0^1 \!\frac{t^{-1/2}(1-t)^{(n-2-\alpha)/3}}
{\bigl(1-\tfrac14 t\bigr)^{(n+1-\alpha)/3}}\,\rmd t
\nonumber\\
&{}=\frac{\sqrt\pi\,\Gamma\bigl(\tfrac{n+1-\alpha}{3}\bigr)}
{\Gamma\bigl(\tfrac{2n+5-2\alpha}{6}\bigr)}\cdot
{}_2F_1\Bigl(\frac{n+1-\alpha}{3},
\frac12;\,\frac{2n+5-2\alpha}{6}\,\Bigl|\,\frac14\Bigr).
\nonumber
\end{align}

As result, the contribution of $I_2$ to $\Mgoth(\alpha,c)$ is
\begin{align}
\frac{\Gamma(\alpha)}{8\pi^3}\Re I_2
=\frac{\Gamma(\alpha)}{24\pi^{5/2}}
\sum_{n=0}^\infty \,\frac{\Gamma^2\bigl(\tfrac{n+1-\alpha}{3}\bigr)}
{\Gamma\bigl(\tfrac{2n+5-2\alpha}{6}\bigr)}&\,{}12^{(n+1-\alpha)/3}
\cos\Bigl(\frac{\pi}{6}(4n+1+2\alpha)\Bigr)\frac{c^n}{n!}
\nonumber\\
&{}\times{}_2F_1\Bigl(\frac{n+1-\alpha}{3},
\frac12;\,\frac{2n+5-2\alpha}{6}\,\Bigl|\,\frac14\Bigr).
\nonumber
\end{align}

Returning to (\ref{eq:Mgoth1}) and applying one of the linear transformations 
of the Gauss hypergeometric function \cite[Eq.(15.8.4)]{DLMF} to our case,
\begin{align}
{}_2F_1\Bigl(a,\frac12;\,1\,\Bigl|\,\frac34\Bigr)
=\frac{\Gamma\bigl(\tfrac12-a\bigr)}{\sqrt\pi\,\Gamma(1-a)}&{}\cdot
{}_2F_1\Bigl(a,\frac12;\,a+\frac12\,\Bigl|\,\frac14\Bigr)
\nonumber\\
&{}+\frac{2^{2a-1}\Gamma\bigl(a-\tfrac12\bigr)}{\sqrt\pi\,\Gamma(a)}\cdot
{}_2F_1\Bigl(1-a,\frac12;\,\frac32-a\,\Bigl|\,\frac14\Bigr),
\nonumber
\end{align}
we get the final result after a few algebraic steps:
\begin{align}
\Mgoth(\alpha,c)&=\frac{\Gamma(\alpha)}{\pi^{3/2}}
\sum_{n=0}^\infty \,\frac{48^{(n-\alpha-2)/3}(-c)^n}
{\Gamma\bigl(\tfrac{2\alpha+7-2n}{6}\bigr)n!}
\cdot{}_2F_1\Bigl(\frac{\alpha+2-n}{3},\frac12;\,
\frac{2\alpha+7-2n}{6}\,\Bigl|\,\frac14\Bigr).
\label{eq:Moments_Ai^4_c}
\end{align}

As a corollary, for $c=0$ we obtain the Mellin transform of the quartic 
product of $\Ai(x)$:
\begin{equation}
\Mgoth[\Ai^4(x)](\alpha)=\frac{\Gamma(\alpha)}{48^{(\alpha+2)/3}\pi^{3/2}
\Gamma\bigl(\tfrac{2\alpha+7}{6}\bigr)}
\cdot{}_2F_1\Bigl(\frac{\alpha+2}{3},\frac12;\,
\frac{2\alpha+7}{6}\,\Bigl|\,\frac14\Bigr).
\label{eq:Moments_Ai^4}
\end{equation}
In particular, the Laurenzi integral (\ref{eq:Laurenzi}) follows immediately 
from (\ref{eq:Moments_Ai^4}) due to the known value 
${}_2F_1\bigl(1,\tfrac12;\,\tfrac32\,\bigl|\,\tfrac14\bigr)=\ln 3$ 
\cite[Eq.(7.3.2.83)]{Prudnikov3_ENG}.

Hypergeometric function values in (\ref{eq:Moments_Ai^4}) may be expressed as 
terminated series for any positive integer $\alpha$. Since the value 
expressions depend on $\alpha\,({\rm mod}\,3)$, we consider the cases 
$\alpha=3m+1$, $\alpha=3m+2$, and $\alpha=3m+3$, where 
$m\in\Nset_0\stackrel{\rm def}{=}\{0\}\cup\Nset$, separately.

For the case $\alpha=3m+1$ the value of the corresponding hypergeometric 
function is presented in \cite[Eq.(8.2.1.22)]{Brychkov_ENG} in terms of 
Jacobi polynomials but, in our view, looks too cumbersome. A simpler 
expression may be found based on (\ref{eq:HyperG_2F1_Int}):
\begin{align}
{}_2F_1\Bigl(m+1,\frac12;\,m+\frac32\,\Bigl|\,a^2\Bigr)
=\frac{\Gamma\bigl(m+\tfrac32\bigr)}{\sqrt\pi\,m!}
\int_0^1 \frac{t^{-1/2}(1-t)^m}{(1-a^2t)^{m+1}}\,\rmd t
=\frac{\bigl(\tfrac32\bigr)_m}{m!}
\int_0^1 \!\frac{(1-t^2)^m}{(1-a^2t^2)^{m+1}}\,\rmd t
\nonumber\\
{}=\frac{\bigl(\tfrac32\bigr)_m}{a^{2m} m!}
\int_0^1 \frac{(a^2-1+1-a^2t^2)^m}{(1-a^2t^2)^{m+1}}\,\rmd t
=\frac{\bigl(\tfrac32\bigr)_m}{a^{2m} m!}
\sum_{k=0}^m \binom{m}{k}(a^2-1)^k\!
\int_0^1 \!\frac{\rmd t}{(1-a^2t^2)^{k+1}}.
\nonumber
\end{align}

Integration by parts yields
$$
\int_0^1 \!\frac{\rmd t}{(1-a^2t^2)^{k+1}}=\frac{(2k)!}{2^{2k}k!^2}
\biggl\{\frac{1}{2a}\ln\frac{1+a}{1-a}+
\sum_{n=1}^k \frac{1}{2n(1-a^2)^n}\cdot\frac{2^{2n}n!^2}{(2n)!}\biggr\}.
$$
Then
\begin{align}
{}_2F_1\Bigl(m+1,\frac12;\,m+\frac32\,\Bigl|\,a^2\Bigr)
=\frac{\bigl(\tfrac32\bigr)_m}{a^{2m}m!}&\,
\biggl\{{}_2F_1\Bigl(-m,\frac12;\,1\,\Bigl|\,1-a^2\Bigr)
\cdot\frac{1}{2a}\ln\frac{1+a}{1-a}
\nonumber\\
&{}+\sum_{k=1}^m \;(-1)^k\binom{m}{k}\!\binom{2k}{k}\!
\sum_{n=1}^k \frac{1}{2n\binom{2n}{n}}
\Bigl(\frac{1-a^2}{4}\Bigr)^{\!k-n}\biggr\}
\label{eq:HyperG_2F1_Rel9}
\end{align}
and the moment integrals are as follows:
\begin{align}
&\Mgoth[\Ai^4(x)](3m+1)
=\frac{(3m)!}{48^m\bigl(\tfrac32\bigr)_m\cdot 24\pi^2}
\cdot{}_2F_1\Bigl(m+1,\frac12;\,m+\frac32\,\Bigl|\,\frac14\Bigr)
\nonumber\\
&{}=\frac{(3m)!}{12^m m!\,24\pi^2}\,
\biggl\{{}_2F_1\Bigl(-m,\frac12;\,1\,\Bigl|\,\frac34\Bigr)\cdot\ln 3
+\sum_{k=1}^m \,(-1)^k\binom{m}{k}\!\binom{2k}{k}\!\sum_{n=1}^k 
\frac{\bigl(\tfrac{3}{16}\bigr)^{k-n}}{2n\binom{2n}{n}}\biggr\}.
\label{eq:Moments_Ai^4_3m+1}
\end{align}

For two other cases the hypergeometric function values may be found with the 
help of the result proven by Vid\=unas \cite{Vidunas3}, namely,
\begin{align}
{}_2F_1\Bigl(-a,\frac12;\,n+2a+\frac32\,\Bigl|\,\frac14\Bigr)
&=\frac{3^{3a}}{2^{2a}}\cdot
\frac{\Gamma\bigl(a+\tfrac12\bigr)\Gamma\bigl(n+2a+\tfrac32\bigr)}
{\sqrt\pi\,\Gamma\bigl(n+3a+\tfrac32\bigr)}\,K(a,n)+{}
\nonumber\\
&{}+(-1)^{n-1}2\cdot 3^{n-2}\cdot
\frac{\Gamma(a+1)\Gamma\bigl(n+2a+\tfrac32\bigr)}
{\Gamma\bigl(a+\tfrac32\bigr)\Gamma(n+2a+1)}\,L(a,n),
\label{eq:Vidunas}
\end{align}
where $a\in\Rset$, $n\in\Zset$ and $K(a,n)$, $L(a,n)$ are terminated 
hypergeometric series:
\begin{gather}
K(a,0)=1,\quad K(a,1)=0,\qquad L(a,0)=0,\quad L(a,1)=1,
\nonumber\\
K(a,n)=\frac{(-1)^n n}{3}
\sum_{\lceil\frac{n}{3}\rceil\le k\le\lfloor\frac{n}{2}\rfloor}
\frac{(k-1)!}{(n-2k)!\,(3k-n)!}\cdot
\frac{\bigl(a+\tfrac12\bigr)_k}{(a+1)_k}\Bigl(\frac{27}{4}\Bigr)^{\!k},
\tag{$n\ge 2$}\\
K(a,-n)=\frac{n}{3}\sum_{0\le k\le\lfloor\frac{n}{3}\rfloor}
\frac{(n-2k-1)!}{(n-3k)!\,k!}\cdot
\frac{(-a)_k}{\bigl(-a+\tfrac12\bigr)_k}\Bigl(-\frac{4}{27}\Bigr)^{\!k},
\tag{$n\ge 1$}\\
L(a,n+1)=\frac{n}{3}\sum_{0\le k\le\lfloor\frac{n}{3}\rfloor}
\frac{(n-2k-1)!}{(n-3k)!\,k!}\cdot
\frac{(a+1)_k}{\bigl(a+\tfrac32\bigr)_k}\Bigl(-\frac{4}{27}\Bigr)^{\!k},
\tag{$n\ge 1$}\\
L(a,1-n)=\frac{(-1)^n n}{3}
\sum_{\lceil\frac{n}{3}\rceil\le k\le\lfloor\frac{n}{2}\rfloor}
\frac{(k-1)!}{(n-2k)!\,(3k-n)!}\cdot
\frac{\bigl(-a-\tfrac12\bigr)_k}{(-a)_k}\Bigl(\frac{27}{4}\Bigr)^{\!k}.
\tag{$n\ge 2$}
\end{gather}

If $\alpha=3m+2$, then $a=-m-\tfrac43$, $n=3m+3$,
\begin{align}
{}_2F_1\Bigl(m+\frac43,\frac12;\,m+\frac{11}6\,\Bigl|\,\frac14\Bigr)
=2^{2/3}\frac{\bigl(\tfrac56\bigr)_{m+1}}{\bigl(\tfrac43\bigr)_m}\,
\biggl\{2\cdot{}_4F_3\biggl(\!\begin{array}{c} 
-\frac{m}{2},-\frac{m+1}{2},\frac16,m+1\smallskip\\ 
\frac13,\frac23,\frac23\end{array}\!\biggl|\,1\biggr)
\nonumber\\
{}-2^{2m-1}(3m+2)(3m+3)\kappa^3\cdot
{}_4F_3\biggl(\!\begin{array}{c} -m,\frac56,\frac{m+2}{2},\frac{m+3}{2}
\smallskip\\ \frac43,\frac43,\frac53\end{array}\!\biggl|\,1\biggr)\biggr\},
\label{eq:HyperG_2F1_14a}
\end{align}
and
\begin{equation}
\Mgoth[\Ai^4(x)](3m+2)=\frac{(3m+1)!}{2^{4m+5}3^{m+1}\pi^2\kappa}
\cdot\frac{1}{2^{2/3}\bigl(\tfrac56\bigr)_{m+1}}\cdot
{}_2F_1\Bigl(m+\frac43,\frac12;\,m+\frac{11}{6}\,\Bigl|\,\frac14\Bigr).
\label{eq:Moments_Ai^4_3m+2}
\end{equation}
If $\alpha=3m+3$, then $a=-m-\tfrac53$, $n=3m+4$,
\begin{align}
{}_2F_1\Bigl(m+\frac53,\frac12;\,m+\frac{13}6\,\Bigl|\,\frac14\Bigr)
=2^{1/3}\frac{\bigl(\tfrac16\bigr)_{m+2}}{\bigl(\tfrac23\bigr)_{m+1}}\,
\biggl\{\frac{2^{2m+3}}{\kappa^3}\cdot
{}_4F_3\biggl(\!\begin{array}{c} -(m+1),\frac16,\frac{m+1}{2},\frac{m+2}{2}
\smallskip\\ \frac13,\frac23,\frac23\end{array}\!\biggl|\,1\biggr)
\nonumber\\
{}-\frac{(3m+3)(3m+4)}{2}\cdot{}_4F_3\biggl(\!
\begin{array}{c} -\frac{m}{2},-\frac{m-1}{2},\frac56,m+2 \smallskip\\ 
\frac43,\frac43,\frac53\end{array}\!\biggl|\,1\biggr)\biggr\},
\label{eq:HyperG_2F1_14b}
\end{align}
and
\begin{equation}
\Mgoth[\Ai^4(x)](3m+3)=\frac{(3m+2)!\,\kappa}{2^{4m+7}3^{m+2}\pi^2}
\cdot\frac{1}{2^{1/3}\bigl(\tfrac16\bigr)_{m+2}}\cdot
{}_2F_1\Bigl(m+\frac53,\frac12;\,m+\frac{13}{6}\,
\Bigl|\,\frac14\Bigr).
\label{eq:Moments_Ai^4_3m+3}
\end{equation}
Here
\begin{equation}
\kappa=3^{1/3}\frac{\Gamma\bigl(\tfrac23\bigr)}{\Gamma\bigl(\tfrac13\bigr)}
=-\frac{\Ai'(0)}{\Ai(0)}\approx 0.7290
\label{eq:kappa}
\end{equation}
is an auxiliary parameter used for brevity.

In particular,
$$
\Mgoth[\Ai^4(x)](2)=\frac{\tfrac23-\kappa^3}{32\pi^2\kappa},\quad
\Mgoth[\Ai^4(x)](3)=\frac{\tfrac{7}{12}-\kappa^3}{64\pi^2\kappa^2},\quad
\Mgoth[\Ai^4(x)](4)=\frac{\tfrac54\,\ln 3-1}{96\pi^2},
$$
where the first two integrals have been found in \cite{Laurenzi}.

\section{The Mellin transform of $(\Ai^3\Bi)(c+x)$ and 
$(\Ai^2\Bi^2)(c+x)$}\label{sec3}
%%%%%%%%%%%%%%%%%%%%%%%%%%%%%%%%%%%%%%%%%%%%%%%%%%%%%%%%%%%%%%%%%%%%%%%%%%%%

It is quite evident that the integral
$$
\Mgoth_1(\alpha,c)=\int_0^\infty x^{\alpha-1}
\bigl\{\Ai^2(c+x)+\rmi\,\Ai(c+x)\Bi(c+x)\bigr\}^2\,\rmd x
$$
is closely connected with the integral $\Mgoth(\alpha,c)$.

Using the Aspnes formula (\ref{eq:Aspnes}) it may be reduced to the first 
component of $\Mgoth(\alpha,c)$:
$$
\Mgoth_1(\alpha,c)
=\frac{\Gamma(\alpha)}{4\pi^3}\rme^{\pi\rmi(\alpha+1)/2}I_1
$$
that leads to the following formulae:
\begin{align}
&\Mgoth[(\Ai^3\Bi)(c+x)](\alpha)=\frac12\Im\Mgoth_1(\alpha,c)
\nonumber\\
&{}=\frac{\Gamma(\alpha)}{2\pi^2}
\sum_{n=0}^\infty \,\Gamma\Bigl(\frac{n+1-\alpha}{3}\Bigr)
12^{(n-\alpha-2)/3}\sin\Bigl(\frac{\pi}{3}(2n+2+\alpha)\Bigr)
\frac{c^n}{n!}\cdot
{}_2F_1\Bigl(\frac{n+1-\alpha}{3},\frac12;\,1\,\Bigl|\,\frac34\Bigr)
\nonumber\\
&{}=\frac{\Gamma(\alpha)}{2\pi}
\sum_{n=0}^\infty \,\frac{12^{(n-\alpha-2)/3}(-c)^n}
{\Gamma\bigl(\tfrac{\alpha+2-n}{3}\bigr) n!}\cdot
{}_2F_1\Bigl(\frac{n+1-\alpha}{3},\frac12;\,1\,\Bigl|\,\frac34\Bigr),
\label{eq:Moments_Ai^3*Bi_c}\\
&\Mgoth[(\Ai^2\Bi^2)(c+x)](\alpha)=\Mgoth(\alpha,c)-\Re\Mgoth_1(\alpha,c)
=\Mgoth[\Ai^4(c+x)](\alpha)
\nonumber\\
&{}-\frac{\Gamma(\alpha)}{\pi^2}
\sum_{n=0}^\infty \,\Gamma\Bigl(\frac{n+1-\alpha}{3}\Bigr)
12^{(n-\alpha-2)/3}\cos\Bigl(\frac{\pi}{3}(2n+2+\alpha)\Bigr)
\frac{c^n}{n!}\cdot
{}_2F_1\Bigl(\frac{n+1-\alpha}{3},\frac12;\,1\,\Bigl|\,\frac34\Bigr).
\label{eq:Moments_Ai^2*Bi^2_c}
\end{align}
Here the convergence conditions are $\alpha>0$ for 
(\ref{eq:Moments_Ai^3*Bi_c}) and $0<\alpha<1$ for 
(\ref{eq:Moments_Ai^2*Bi^2_c}).

In particular, for $c=0$ we have
\begin{align}
\Mgoth[(\Ai^3\Bi)(x)](\alpha)&=\frac{\Gamma(\alpha)}
{12^{(\alpha+2)/3}2\pi\Gamma\bigl(\tfrac{\alpha+2}{3}\bigr)}\cdot
{}_2F_1\Bigl(\frac{1-\alpha}{3},\frac12;\,1\,\Bigl|\,\frac34\Bigr),
\label{eq:Moments_Ai^3*Bi}\\
\Mgoth[(\Ai^2\Bi^2)(x)](\alpha)
&=\Mgoth[\Ai^4(x)](\alpha)
+\frac{\Gamma(\alpha)\Gamma\bigl(\tfrac{1-\alpha}{3}\bigr)}
{12^{(\alpha+2)/3}\pi^2}\,\cos\frac{\pi(1-\alpha)}{3}\cdot
{}_2F_1\Bigl(\frac{1-\alpha}{3},\,\frac12;\,1\,\Bigl|\,\frac34\Bigr).
\label{eq:Moments_Ai^2*Bi^2}
\end{align}

For any $\alpha\in\Nset$ the integral $\Mgoth[(\Ai^3\Bi)(x)](\alpha)$ has a 
simpler form because its hypergeometric factor may be reduced to terminated 
series. The approach here is absolutely the same as for 
$\Mgoth[\Ai^4(x)](\alpha)$ above and we provide the resulting expressions only:
\begin{align}
&\Mgoth[(\Ai^3\Bi)(x)](3m+1)=\frac{(3m)!}{12^m m!\,24\pi}\cdot
{}_2F_1\Bigl(-m,\frac12;\,1\,\Bigl|\,\frac34\Bigr),
\label{eq:Moments_Ai^3*Bi_3m+1}\\
&\Mgoth[(\Ai^3\Bi)(x)](3m+2)
\nonumber\\
&{}=\frac{1}{12^m\bigl(\tfrac43\bigr)_m 64\pi^2\sqrt 3}\,
\biggl\{(3m+3)!\,\kappa^2\cdot
{}_4F_3\biggl(\!\begin{array}{c} -m,\frac56,\frac{m+2}{2},\frac{m+3}{2}
\smallskip\\ \frac43,\frac43,\frac53\end{array}\!\biggl|\,1\biggr)+{}
\nonumber\\
&\hspace{40mm}{}+\frac{(3m+1)!}{2^{2m-1}\kappa}\cdot{}_4F_3\biggl(\!
\begin{array}{c} -\frac{m}{2},-\frac{m+1}{2},\frac16,m+1\smallskip\\ 
\frac13,\frac23,\frac23\end{array}\!\biggl|\,1\biggr)\biggr\},
\label{eq:Moments_Ai^3*Bi_3m+2}\\
&\Mgoth[(\Ai^3\Bi)(x)](3m+3)
\nonumber\\
&{}=\frac{1}{12^m\bigl(\tfrac53\bigr)_m 32\pi^2\sqrt 3}\,
\biggl\{\frac{(3m+2)!}{\kappa^2}\cdot
{}_4F_3\biggl(\!\begin{array}{c} -(m+1),\frac16,\frac{m+1}{2},\frac{m+2}{2}
\smallskip\\ \frac13,\frac23,\frac23\end{array}\!\biggl|\,1\biggr)+{}
\nonumber\\
&\hspace{40mm}{}+\frac{(3m+4)!}{2^{2m+5}}\,\kappa\cdot
{}_4F_3\biggl(\!\begin{array}{c} -\frac{m}{2},-\frac{m-1}{2},\frac56,m+2 
\smallskip\\ \frac43,\frac43,\frac53\end{array}\!\biggl|\,1\biggr)\biggr\},
\label{eq:Moments_Ai^3*Bi_3m+3}
\end{align}
where $m\in\Nset_0$ and $\kappa$ is defined by (\ref{eq:kappa}).

It is interesting to note the following correspondence between moment 
integrals of $\Ai^4(x)$ and $\Ai^3(x)\Bi(x)$:
\begin{align}
\Mgoth[\Ai^4(x)](3m+1)&=\frac{p_1\ln 3+q_1}{\pi^2}, &
\Mgoth[(\Ai^3\Bi)(x)](3m+1)&=\frac{p_1}{\pi},
\label{eq:Corres1}\\
\Mgoth[\Ai^4(x)](3m+2)&=\frac{p_2-q_2\kappa^3}{\pi^2\kappa}, &
\Mgoth[(\Ai^3\Bi)(x)](3m+2)&=\frac{(p_2+2q_2\kappa^3)\sqrt 3}{2\pi^2\kappa},
\label{eq:Corres2}\\
\Mgoth[\Ai^4(x)](3m+3)&=\frac{p_3-2q_3\kappa^3}{3\pi^2\kappa^2}, &
\Mgoth[(\Ai^3\Bi)(x)](3m+3)&=\frac{(p_3+q_3\kappa^3)\sqrt 3}{2\pi^2\kappa^2},
\label{eq:Corres3}
\end{align}
where $p_k$, $q_k$ ($k=1,2,3$) are rational numbers depending on $m$.

A similar correspondence exists for any other value of $\alpha$. Namely, let 
$\alpha=3m+\beta$, where $m\in\Nset_0$ and $\beta\in(0,3]$. Then equations 
(\ref{eq:Moments_Ai^4}) and (\ref{eq:Moments_Ai^3*Bi}) define two sequences 
on~$m$:
\begin{gather}
\Mgoth[\Ai^4(x)](3m+\beta)=\frac{\Gamma(3m+\beta)}
{12^m 48^b\pi^{3/2}\Gamma\bigl(b+\tfrac12\bigr)}\cdot y_m(\beta),
\label{eq:Moments_Ai^4_beta}\\
\Mgoth[(\Ai^3\Bi)(x)](3m+\beta)
=\frac{\Gamma(3m+\beta)}{12^{m+b}2\pi\Gamma(b)}\cdot z_m(\beta),
\label{eq:Moments_Ai^3Bi_beta}
\end{gather}
where $b=(\beta+2)/3$ is an auxiliary parameter used for brevity and
\begin{gather}
y_m(\beta)=\frac{1}{4^m\bigl(b+\tfrac12\bigr)_m}\cdot
{}_2F_1\Bigl(m+b,\frac12;\,m+b+\frac12\,\Bigl|\,\frac14\Bigr),
\label{eq:Seq_ym}\\
z_m(\beta)=\frac{1}{(b)_m}\cdot
{}_2F_1\Bigl(1-b-m,\frac12;\,1\,\Bigl|\,\frac34\Bigr).
\label{eq:Seq_zm}
\end{gather}
Applying well-known recurrence relations for the Gauss hypergeometric 
functions (see, for example, Eqs.(7.3.1.11) and (7.3.1.19) in 
\cite{Prudnikov3_ENG}):
$$
\begin{array}{cc}
f_m={}_2F_1\bigl(m+b,\tfrac12;\,m+b+\tfrac12\,\bigl|\,\tfrac14\bigr){:} &
(m+b)^2f_{m+1}-\bigl\{(m+b)^2-\tfrac14\bigr\}(5f_m-4f_{m-1})=0,
\medskip\\
f_m={}_2F_1\bigl(1-b-m,\tfrac12;\,1\,\bigl|\,\tfrac34\bigr){:} &
4(m+b)f_{m+1}-5\bigl(m+b-\tfrac12\bigr)f_m+(m+b-1)f_{m-1}=0,
\end{array}
$$
it is easy to check that both sequences, $y_m(\beta)$ and $z_m(\beta)$, 
satisfy one and the same recurrence relation 
\begin{equation}
4(m+b)^2f_{m+1}-5\bigl(m+b-\tfrac12\bigr)f_m+f_{m-1}=0
\label{eq:BothIntegRecur}
\end{equation}
(see also Eq.(34) in \cite{Laurenzi}).

A general solution of (\ref{eq:BothIntegRecur}) is uniquely defined by initial 
conditions $f_0$, $f_1$ and may be written as follows:
\begin{equation}
f_m(b\,|\,f_0,f_1)=f_0P_m(b)+f_1Q_m(b),
\label{eq:BothIntegRecurSolution}
\end{equation}
where
\begin{align}
P_m(b)=f_m(b\,|\,1,0)&=\Bigl\{1,\,0,\,-\frac{1}{4(1+b)^2},
-\frac{5\bigl(\tfrac32+b\bigr)}{4^2(1+b)^2(2+b)^2},\,\ldots\Bigr\},
\label{eq:Seq_Pm}\\
Q_m(b)=f_m(b\,|\,0,1)&=\Bigl\{0,\,1,\,
\frac{5\bigl(\tfrac12+b\bigr)}{4(1+b)^2},
\frac{5^2\bigl(\tfrac12+b\bigr)\bigl(\tfrac32+b\bigr)}{4^2(1+b)^2(2+b)^2}
-\frac{1}{4(2+b)^2},\,\ldots\Bigr\}.
\label{eq:Seq_Qm}
\end{align}
Returning to the integrals $\Mgoth[\Ai^4(x)](3m+\beta)$ and 
$\Mgoth[(\Ai^3\Bi)(x)](3m+\beta)$ we conclude that the dependence of 
the factors
\begin{equation}
y_m(\beta)=y_0(\beta)P_m(b)+y_1(\beta)Q_m(b),\qquad 
z_m(\beta)=z_0(\beta)P_m(b)+z_1(\beta)Q_m(b)
\label{eq:Solution_ym_zm}
\end{equation}
on $m$ is determined by the same terms of $P_m(b)$ and $Q_m(b)$.
In particular, if $\beta$ is a rational, then both sequences $P_m(b)$ and 
$Q_m(b)$ contain rational numbers only. (We recall that $b=(\beta+2)/3$.)

If $\alpha=3m+\tfrac52$, then $\Mgoth[\Ai^4(x)](\alpha)$ and 
$\Mgoth[(\Ai^3\Bi)(x)](\alpha)$ are reduced to linear combinations of special 
values of complete elliptical integrals because corresponding hypergeometric 
factors are known \cite{Prudnikov3_ENG}:
\begin{gather}
\Mgoth[\Ai^4(x)]\Bigl(3m+\frac52\Bigr)
=\frac{(6m+4)!}{2^{8m+10}3^{m+3/2}(3m+2)!\,\pi}\cdot y_m\Bigl(\frac52\Bigr),
\label{eq:Moments_Ai^4_3m+2.5}\\
y_0\Bigl(\frac52\Bigr)=\frac{16}{\pi}
\Bigl\{{\bf K}\Bigl(\frac12\Bigr)-{\bf E}\Bigl(\frac12\Bigr)\Bigr\},\qquad 
y_1\Bigl(\frac52\Bigr)=\frac{8}{9\pi}
\Bigl\{9{\bf K}\Bigl(\frac12\Bigr)-10{\bf E}\Bigl(\frac12\Bigr)\Bigr\},
\label{eq:Seq_y0_y1}\\
\Mgoth[(\Ai^3\Bi)(x)]\Bigl(3m+\frac52\Bigr)
=\frac{(6m+4)!}{2^{8m+7}3^{m+3/2}(3m+2)!\,\pi}\cdot z_m\Bigl(\frac52\Bigr),
\label{eq:Moments_Ai^3Bi_3m+2.5}\\
z_0\Bigl(\frac52\Bigr)=\frac{2}{\pi}{\bf E}\Bigl(\frac{\sqrt 3}2\Bigr),\qquad
z_1\Bigl(\frac52\Bigr)=\frac{1}{6\pi}
\Bigl\{10{\bf E}\Bigl(\frac{\sqrt 3}2\Bigr)
-{\bf K}\Bigl(\frac{\sqrt 3}2\Bigr)\Bigr\}.
\label{eq:Seq_z0_z1}
\end{gather}
In particular,
\begin{align}
\Mgoth[\Ai^4(x)]\Bigl(\frac52\Bigr)
&=\frac{1}{32\pi^2\sqrt{3}}
\Bigl\{{\bf K}\Bigl(\frac12\Bigr)-{\bf E}\Bigl(\frac12\Bigr)\Bigr\},
\label{eq:Moments_Ai^4_2.5}\\
\Mgoth[(\Ai^3\Bi)(x)]\Bigl(\frac52\Bigr)
&=\frac{1}{16\pi^2\sqrt{3}}{\bf E}\Bigl(\frac{\sqrt 3}2\Bigr).
\label{eq:Moments_Ai^3*Bi_2.5}
\end{align}

\begin{figure}[t]
\centerline{%
\includegraphics[width=0.45\textwidth,keepaspectratio]{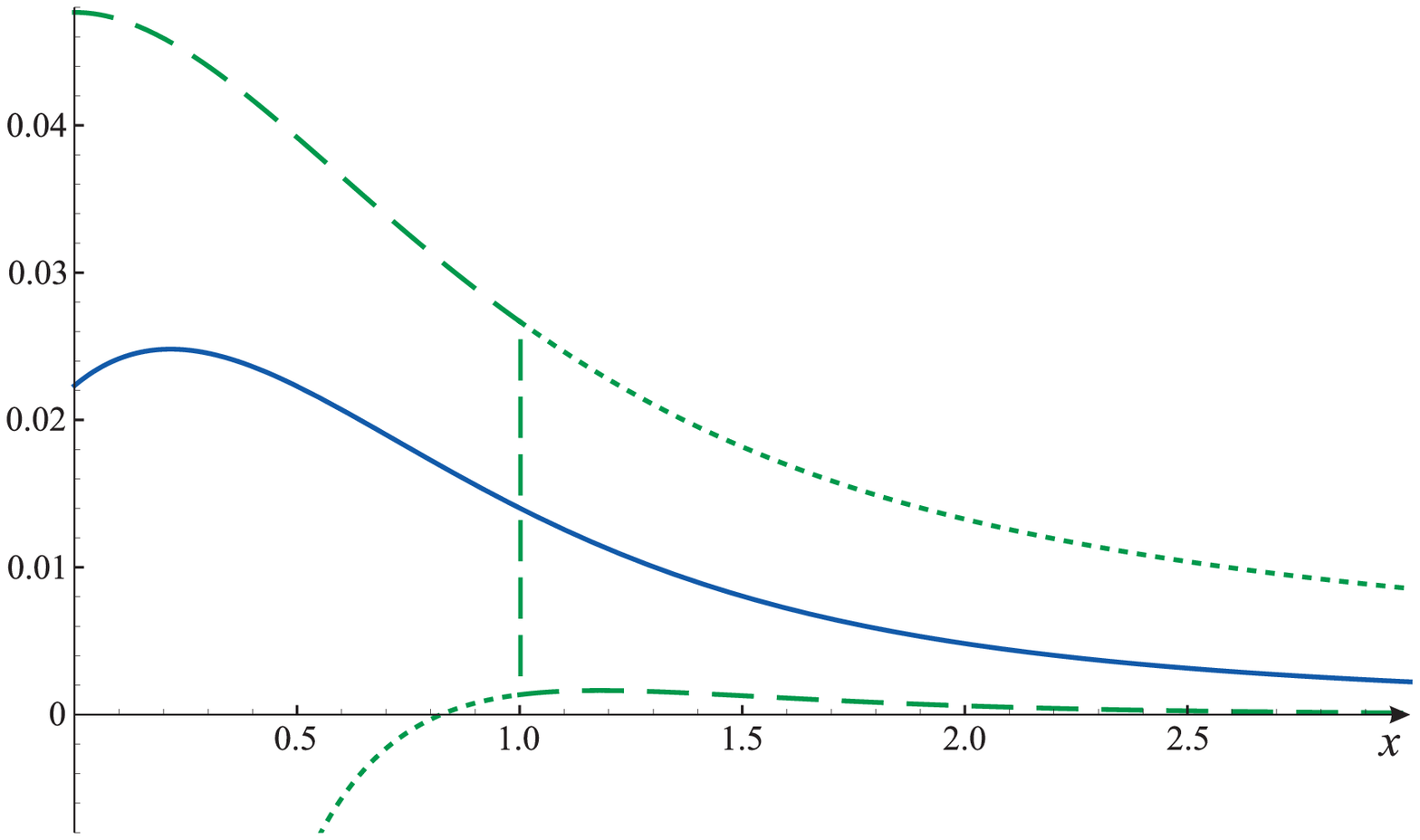}
~~
\includegraphics[width=0.45\textwidth,keepaspectratio]{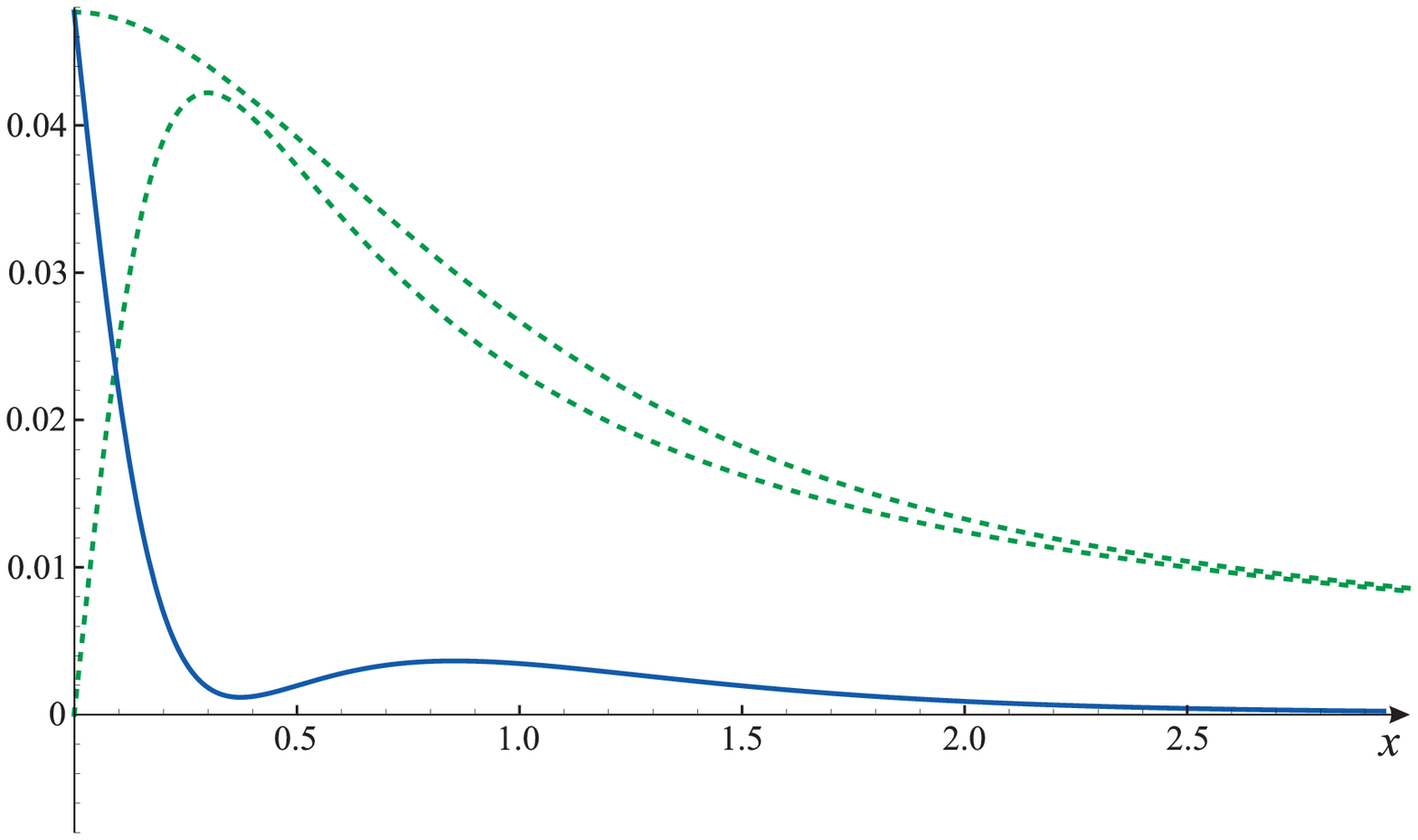}
}
\noindent
\caption[]{\small Left: integrands of Eq.(\ref{eq:Ai2Bi2_b}) (dashed 
line) and Eq.(\ref{eq:Ai2Bi2_c}) (continuous line). Right: the integrand of 
Eq.(\ref{eq:Ai2Bi2_d}) (continuous line) and its components\par}
\label{fig:Ai2Bi2_Reid}
\end{figure}

As for the integral $\Mgoth[(\Ai^2\Bi^2)(x)](\alpha)$, we do not know any 
value of $\alpha$ for which the hypergeometric factor on the right 
side of (\ref{eq:Moments_Ai^2*Bi^2}) may be expressed in a simpler form. 
Only for the case of upper limit of the convergence interval, a kind of 
simplification is possible. Namely, since
\begin{align}
\Ai^2(x)\Bi^2(x)&=\frac{1}{4\pi^2 x}+{\cal O}\Bigl(\frac{1}{x^4}\Bigr)
\qquad (x\to\infty),
\nonumber\\
\Mgoth[\Ai^2(x)\Bi^2(x)](\alpha)&=\frac{1}{4\pi^2(1-\alpha)}
+\frac{4\gamma+12\ln 2-\ln 3}{24\pi^2}+{\cal O}(\alpha-1)\qquad (\alpha\to 1-),
\nonumber
\end{align}
where $\gamma$ is the Euler constant, then utilizing the divergent component:
\begin{align}
\int_0^a &x^{\alpha-1}\Ai^2(x)\Bi^2(x)\,\rmd x
+\int_a^\infty x^{\alpha-1}\Bigl(\Ai^2(x)\Bi^2(x)
-\frac{1}{4\pi^2(x+b)}\Bigr)\,\rmd x
\nonumber\\
&{}=\Mgoth[\Ai^2(x)\Bi^2(x)](\alpha)
-\frac{1}{4\pi^2}\int_a^\infty x^{\alpha-2}\,\rmd x
+\frac{1}{4\pi^2}\int_a^\infty x^{\alpha-1}
\Bigl(\frac{1}{x}-\frac{1}{x+b}\Bigr)\,\rmd x
\nonumber\\
&{}=\frac{4\gamma+12\ln 2-\ln 3}{24\pi^2}+\frac{\ln (a+b)}{4\pi^2}
+{\cal O}(\alpha-1),
\nonumber
\end{align}
where $a$ and $b$ are real parameters, $a+b>0$, one can obtain
\begin{align}
\int_0^a \Ai^2(x)\Bi^2(x)\,\rmd x
+\int_a^\infty \Bigl(\Ai^2(x)\Bi^2(x)-\frac{1}{4\pi^2(x+b)}\Bigr)\,\rmd x
\nonumber\\
{}=\frac{4\gamma+12\ln 2-\ln 3}{24\pi^2}+\frac{\ln (a+b)}{4\pi^2}.
\label{eq:Ai2Bi2_a}
\end{align}
In particular,
\begin{gather}
\int_0^1 \Ai^2(x)\Bi^2(x)\,\rmd x
+\int_1^\infty \Bigl(\Ai^2(x)\Bi^2(x)-\frac{1}{4\pi^2x}\Bigr)\,\rmd x
=\frac{4\gamma+12\ln 2-\ln 3}{24\pi^2},
\label{eq:Ai2Bi2_b}\\
\int_0^\infty \Bigl(\Ai^2(x)\Bi^2(x)-\frac{1}{4\pi^2(x+1)}\Bigr)\,\rmd x
=\frac{4\gamma+12\ln 2-\ln 3}{24\pi^2}.
\label{eq:Ai2Bi2_c}
\end{gather}
The first formula is known due to Reid \cite[Eq.(3.14)]{Reid3}. Changing a 
compensating term in the integrand it is easy to find other similar formulae. 
For example,
\begin{equation}
\int_0^\infty 
\Bigl(\Ai^2(x)\Bi^2(x)-\frac{4x}{\pi^2(16x^2+3^{1/3})}\Bigr)\,\rmd x
=\frac{\gamma}{6\pi^2}.
%%% проверено в Mathematica %%%%%%%%%%%%%%%%%%%%%%%%%%%%%%%%%%%%%%%%%%%%%%
\label{eq:Ai2Bi2_d}
\end{equation}
Plots of integrands of all three last equations are shown in 
Fig.~\ref{fig:Ai2Bi2_Reid}.

\section{Concluding remarks}\label{sec4}
%%%%%%%%%%%%%%%%%%%%%%%%%%%%%%%%%%%%%%%%%%%%%%%%%%%%%%%%%%%%%%%%%%%%%%%%%%%%

The above results help to find the Mellin transform of some other products 
of the Airy functions, for example,
$$
\int_0^\infty x^{\alpha-1}\Ai^4(c-x)\,\rmd x,
$$
where $0<\alpha<1$. Besides, the approach with the change of variables 
$(s,t)\to(u,v)$ may be used to obtain similar results for quadratic products 
of Airy functions, and we are planning to publish them in future.

Quartic products of the Airy functions are also a source of various relations 
containing hypergeometric functions with special values. Namely, if one 
considers the integrand above to be a product of two factors and expands one 
of them into a Taylor series, then integrating term by term one can obtain 
a solution as a power series in~$c$. Selecting the factor for the Taylor 
series expansion in two different ways we will have two different series 
expressions for one and the same solution. Finally, equating coefficients of 
equal powers of both series leads to a relation between hypergeometric 
functions. The simplest examples of this kind are
\begin{align}
{}_2F_1\Bigl(a,\frac12;\,3a\,\Bigl|\,\frac34\Bigr)
&=\frac{2^{2a+1}\sqrt\pi\,\Gamma(3a)}
{3^{3a}\Gamma\bigl(a+\tfrac12\bigr)\Gamma(2a)},
\label{eq:HyperG_2F1_a}\\
{}_2F_1\Bigl(a+\frac12,\frac12;\,3a\,\Bigl|\,\frac34\Bigr)
&=\frac{2\Gamma\bigl(a+\tfrac56\bigr)\Gamma(3a)}
{\Gamma\bigl(a+\tfrac13\bigr)\Gamma\bigl(3a+\tfrac12\bigr)},
\label{eq:HyperG_2F1_b}\\
{}_2F_1\Bigl(3a-\frac54,\frac34-a;\,2a\,\Bigl|\,\frac14\Bigr)
&=\frac{2^{4a-1}\sqrt{\pi}\,\Gamma(2a)}
{3^{3a-3/4}\Gamma\bigl(\tfrac23\bigr)\Gamma\bigl(2a-\tfrac16\bigr)}.
\label{eq:HyperG_2F1_c}
\end{align}
All three formulae are already known: the first one is Eq.(7.3.9.35) 
in \cite{Prudnikov3_ENG}, two others are equivalent to 
Eq.(A.22) in \cite[page~98]{Ebisu} due to the Euler transformation and 
Eq.(ix) in \cite[page~35]{Ebisu}, respectively. This part of our 
research is ongoing.

\section*{Acknowledgements}
%%%%%%%%%%%%%%%%%%%%%%%%%%%%%%%%%%%%%%%%%%%%%%%%%%%%%%%%%%%%%%%%%%%%%%%%%%%%

The authors are grateful to Prof. B.J.~Laurenzi for his valuable comments 
that helped to improve the manuscript.

%\bibliographystyle{unsrt}
%\bibliographystyle{iopart-num}
%\bibliography{all}

\end{document}